# Defining Least Community as a Homogeneous Group in Complex Networks

Bin Jiang and Ding Ma

Faculty of Engineering and Sustainable Development, Division of Geomatics
University of Gävle, SE-801 76 Gävle, Sweden
Email: bin.jiang|ding.ma@hig.se

**Highlights:**
- A new community detection algorithm inspired by the head/tail breaks.
- A new way of thinking for community detection or classification in general.
- Far more small communities than large ones in complex networks.
- Simple networks like mechanical watches, while complex networks like human brains.
- Empirical evidence on power laws of the detected communities.

**Abstract**
This paper introduces a new concept of least community that is as homogeneous as a random graph, and develops a new community detection algorithm from the perspective of homogeneity or heterogeneity. Based on this concept, we adopt head/tail breaks – a newly developed classification scheme for data with a heavy-tailed distribution – and rely on edge betweenness given its heavy-tailed distribution to iteratively partition a network into many heterogeneous and homogeneous communities. Surprisingly, the derived communities for any self-organized and/or self-evolved large networks demonstrate very striking power laws, implying that there are far more small communities than large ones. This notion of far more small things than large ones constitutes a new fundamental way of thinking for community detection.

**Keywords:** head/tail breaks, ht-index, scaling, k-means, natural breaks, and classification

## I. Introduction

A network or graph is simply a set of vertices or nodes joined by edges or links. Sometimes, the edges are given a direction or weight. This paper considers only a binary network, neither a direction nor a weight for all of the edges. A network may be as large as having thousands, millions and even billions of nodes and edges. Large-scale real-world networks are of primary interest because of their large sizes and structural complexity, herein complex networks. Complex networks could be social such as friendships and collaborations, biological such as protein interactions and food webs, technological such as the Internet and streets, and informational such as the World Wide Web (e.g., Cohen and Havlin 2010, Newman 2010). Large networks are unlikely to be regular, such as lattices (e.g., crystal in reality) in which each node has the same number of edges, or random (e.g., gas in a container), in which every pair of nodes has the same probability of being linked (both regular and random graphs with homogeneous structures); instead, they are something in between regular and random. Subsequently, real-world networks differ fundamentally from their random counterparts in that they display a very significant heterogeneity. This heterogeneity implies that complex networks consist of many different relatively independent compartments, leading to the notion of community or community structure.

A community, also called a module or a cluster, is loosely defined as a subset of vertices with many inside edges and a few outside edges. In other words, vertices within a community are densely connected, whereas connections or edges between communities are sparser. This definition sounds very intuitive and straightforward; however, it is hardly operable for community detection, in particular for large or complex networks. This paper introduces a new concept of least community as a homogeneous group – as homogeneous as a random graph. Based on this concept, a heterogeneous network is partitioned into many homogeneous communities by referencing its random graph. The random graph is used as a reference because it is considered homogeneous enough and its edges are imposed with the same probability, or it contains only one community. Considering a network as a set of edges characterized by the measure edge betweenness (Girvan and Newman 2002), the issue of community detection becomes that of classification, i.e., classifying all edges into different homogeneous groups as homogeneous as a random graph or, more specifically, into inside and outside edges.

The classification relies on edge betweenness to determine different classes or communities. The edge betweenness of real-world networks demonstrates a heavy tail distribution, indicating that conventional methods such as k-means (MacQueen 1967) and natural breaks (Jenks 1967) could not effectively derive the classes that reflect the underlying scaling pattern. These conventional methods use the mean or the average to characterize individual classes, but the edge betweenness is right skewed or scale free. Given the circumstance, head/tail breaks, a newly developed classification scheme (Jiang 2013), is more appropriate and effective for data with a heavy tail distribution. Head/tail breaks partitions all the edges into the head (those edges with betweenness greater than the mean) and the tail (those edges with betweenness less than the mean), and recursively continues the partition process until the head percentage is as large as that of the random graph (c.f., the next section for illustrations). This ending condition implies that the head and tail are well balanced, and the derived classes or communities are homogeneous enough. During the recursive partition process, some heterogeneous communities are identified as well. Eventually, both homogenous and heterogeneous communities are derived at different coarse-graining levels. The central argument of this paper is that any self-organized and/or naturally evolved real world network contains far more small communities than large ones, or its communities exhibit a power law or heavy-tailed distribution in general.

Community structure or community detection has received disproportionate attention in the past years, largely because of the availability of rich data from the Internet and social media, and its far-reaching implications for a variety of disciplines (e.g., Fortunato 2010, Newman 2004). Communities could be social groupings in a social network based on interest, related papers in a citation network, related researchers in a collaboration network, functional groupings in a metabolic network such as cycles and pathways, and web pages in a website on the same or similar topics. Both community structure and community detection return large amounts of hits in Google Scholar. Despite the literature on the topic having a long history that dates back to the 1920s (Rice 1927), a vast majority of the studies was conducted in the past decade, in particular since the seminal work of Girvan and Newman (2002). The algorithm developed in this paper brings new insights into community detection or classification in general.

The next section presents the new algorithm and illustrates how a network may be partitioned into many communities, both homogeneous and heterogeneous. Section III reports on our experiments by applying the community detection algorithm to many complex networks, including social, biological, technological, and informational. Finally, Section IV concludes the paper with further discussions.

**II. The new community detection algorithm based on head/tail breaks**
This section illustrates the new community detection algorithm based on head/tail breaks using two sample networks. We start with a fictive social network consisting of 12 vertices and 20 edges (Figure 1). Intuitively, the fictive network contains three communities of sizes 5, 4, and 3. We first create a random network that is the counterpart of the fictive network with the same number of vertices and edges (Panel C of Figure 1). The edge betweenness of the fictive network is very heterogeneous, with a maximum-to-minimum ratio of 19.9, whereas that of the random network is relatively homogeneous, with a maximum-to-minimum ratio of 3.5. As reflected in the corresponding rank-size plots (Panels D and E), the heterogeneity and homogeneity are indicated respectively by the steep and flat distribution curves. The red dots of the curves constitute the head, which consists of edges (or outside edges) with edge betweenness greater than the mean, whereas the remaining blue dots represent the tail for edges (or inside edges) with betweenness less than the mean. The heterogeneity and homogeneity are also reflected in the low and high head/tail ratios: 35/65 attributable to seven edges in the head and 40/60 attributable to eight edges in the head. This low head/tail ratio of 35/65 shows that, potentially, the fictive network contains some communities. Eventually, all seven outside edges are removed, leading to the four communities shown in Panel B.

This example is simple enough for illustrating the basic principle of the community detection algorithm. Formally, the algorithm can be described using the following recursive function; a program is available at https://github.com/digmaa/HeadTailCommunityDetection.

```
Recursive Function Head/tailCommunity (network, head)
// network is to be portioned into communities, head is the head percentage
// of the equivalent random graph)
      Extract all subnetworks of the input network;
// subnetworks each of which contains more than one node
  Foreach subnetwork
      Calculate edge betweenness for each edge;
      Calculate mean betweenness of all betweenness;
      Calculate head percentage in this subnetwork;
      // the number of edges >= mean betweenness value divided by
      // the number of edges of this subnetwork
         If (head percentage >= head)
             Add all subnetwork edges into EdgeList;
         Else
             Call Function Head/tailCommunity (subnetwork, head);
  Return EdgeList;
End Function
```

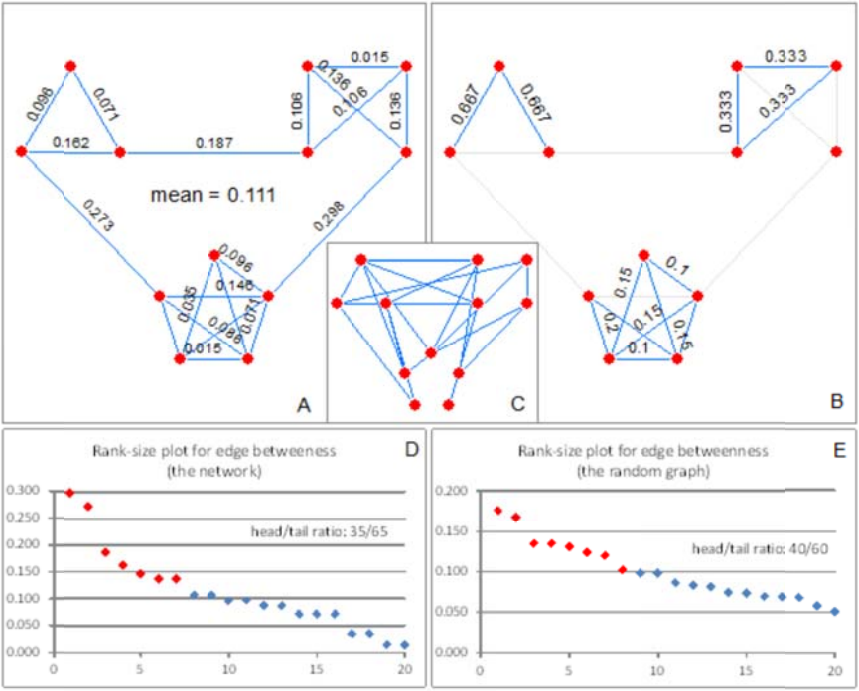

Figure 1: (Color online) Illustration of the community detection algorithm
(Note: A fictive social network (Panel A) with 12 vertices linked by 20 edges, its four detected communities with sizes 5, 3, 3, and 1, shown in Panel B, in which the removed edges are shown in gray, and its equivalent random graph (Panel C); the edge betweenness of the network is heterogeneous with a head/tail ratio of 35/65 (Panel D), whereas that of the random network is homogeneous with a head/tail ratio of 40/60 (Panel E). The random graph's head percentage is just a reference to determine whether the fictive network is heterogeneous, i.e., the fictive network's head is smaller than that of the random graph. The actual partition relies on the network's mean or its head percentage rather than that of the random graph.)

For large networks, the process of removing outside edges should continue iteratively or recursively for the head until the ending condition is met, i.e., the head percentage is greater than or equal to that of the random graph. The community detection process ends up with homogeneous communities that are as homogeneous as the random graph – the ultimate goal for the community detection. Before the last iteration, all previous iterations end up with some heterogeneous communities. To illustrate, we apply the algorithm to the Zachary (1977) Karate Club network that has been widely studied for community detection (e.g., Fortunato 2010, Newman 2004). The network consists of 34 vertices and 78 edges shown in Figure 2, with outside edges marked in gray. The partition process goes step-by-step iteratively. For example, the first iteration ends up with three communities of sizes 28, 5, and 1, two of which are heterogeneous. In the second iteration, the two heterogeneous communities are further partitioned into heterogeneous and homogeneous ones. We can remark that the two known communities of the club reflect pretty well in the second and third iteration results (Figure 2). Eventually, the last iteration ends up with 11 homogeneous communities of sizes 10, 6, 5, 3, 2, 2, 2, 1, 1, 1, and 1. Putting both homogeneous and heterogeneous together results in 14 communities of sizes 28, 15, 10, 6, 5, 5, 3, 2, 2, 2, 1, 1, 1, and 1. Apparently, there are far more small communities than large ones.

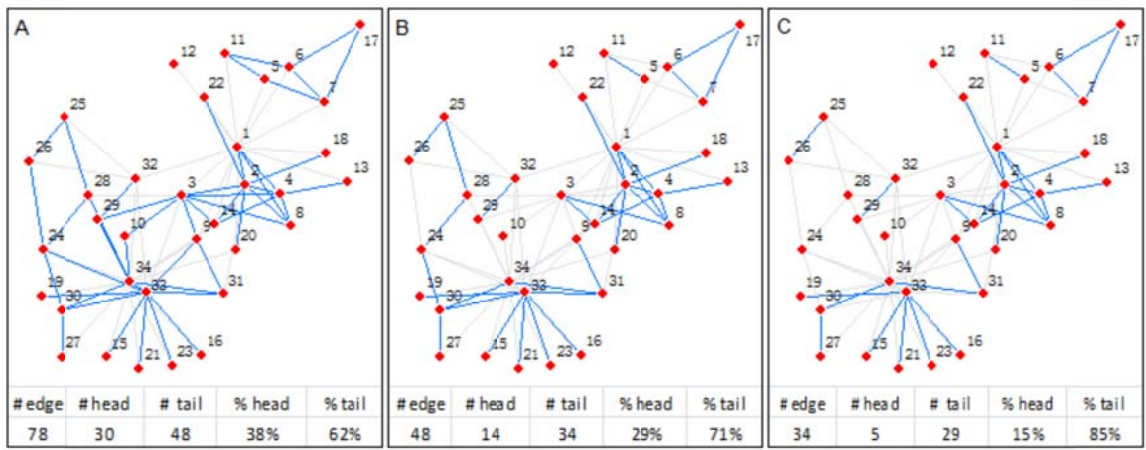

Figure 2: (Color online) Head/tail breaks for detecting communities from the club network
(Note: The network is clearly split along one diagonal direction into two or three major parts with three iterations (Panels A, B, and C). There are far more small communities than large ones: 28, 15, 10, 6, 5, 5, 3, 2, 2, 2, 1, 1, 1, and 1; the numbers below each panel indicate how the edges were split recursively into the head and tail or, equivalently, the outside and inside edges. The scaling pattern of far more inside edges than outside edges remains persistent for all the iterations.)

Most heterogeneous and homogeneous communities have nested relationships, constituting a scaling hierarchy. For example, the largest community of size 28 contains communities of sizes 15, 10, 2, and 1, out of which the community of size 15 further contains communities of sizes 6, 5, 2, 1, and 1. Let us imagine the club network as the body of an alien partitioned into three organs (Panel A), seven tissues (Panel B), and 11 cells (Panel C) with some single-cell organs or tissues. This imagination helps us to see the power of heterogeneous and homogeneous communities in understanding the functional modules of complex networks. The club network example shows also how our definition of community converges to the loose definition that can be slightly reformatted as "far more the inside edges than the outside ones". Indeed, this statement is true at each iteration – the inside edges (or the number in the tail) are always far more than the outside edges (or the number in the head).

## III. Experiments with more real-world networks

We apply the new community detection algorithm to eight complex networks, including social, biological, technological, and informational, to examine whether (1) there are far more small betweenness edges than large ones, and (2) there are far more small communities than large ones. The network data come from various sources: the Brightkite and Gowalla networks from Cho, Myers, and Leskovec (2011) and Jiang and Miao (2014); the scientific, protein, Internet, and streets (Paris) networks from Zhou and Mondragón (2007) and Jiang et al. (2014); and the Erdös and WWW networks from the Pajek datasets at http://vlado.fmf.uni-lj.si/pub/networks/data/. Note that, due to limit for computing the edge betweenness, the Brightkite, Gowalla, and WWW networks are sampled from larger networks.

Table 1: Statistics and results of community detection for eight complex networks
(Note: # node = number of nodes, degree = average degree of connections, # edge = number of edges, ht-edge = ht-index for edge betweenness, iteration = number of iterations to derive communities, # hcom = number of homogeneous communities, ht-hcom = ht-index for homogeneous communities, max = maximum size of communities, min = minimum size of communities, which is always 1, h/t = head/tail ratio of the corresponding random graphs with respect to their edge betweenness, NA = not available)

|  | # node | degree | # edge | ht-edge | iteration | # hcom | ht-hcom | max | min | h/t |
|---|---|---|---|---|---|---|---|---|---|---|
| Brightkite | 288 | 78.6 | 11324 | 8 | 10 | 172 | 3 | 27 | 1 | 25/75 |
| random | 288 | 77.7 | 11188 | NA | NA | NA | NA | NA | NA | 48/52 |
| scalefree | 282 | 75.1 | 10593 | 6 | 14 | 258 | 3 | 14 | 1 | 24/76 |
| smallworld | 288 | 78.0 | 11232 | 10 | 15 | 171 | 5 | 11 | 1 | 43/57 |
| Gowalla | 536 | 112.4 | 32794 | 6 | 21 | 371 | 4 | 12 | 1 | 26/74 |
| random | 536 | 112.4 | 30473 | NA | NA | NA | NA | NA | NA | 49/51 |
| scalefree | 536 | 112.4 | 30122 | 1 | 1 | NA | NA | NA | NA | 49/51 |
| smallworld | 536 | 112.0 | 30016 | 2 | 20 | 316 | 5 | 11 | 1 | 42/58 |
| Scientific | 12271 | 6.5 | 39967 | 10 | 10 | 4799 | 8 | 80 | 1 | 21/79 |
| random | 12245 | 6.5 | 39698 | NA | NA | NA | NA | NA | NA | 43/57 |
| scalefree | 9295 | 8.6 | 40139 | 10 | 12 | 7195 | 5 | 14 | 1 | 34/66 |
| smallworld | 12263 | 6.0 | 36813 | 1 | NA | NA | NA | NA | NA | 44/56 |
| Erdös | 6927 | 3.4 | 11301 | 7 | 9 | 1681 | 6 | 143 | 1 | 17/83 |
| random | 6612 | 3.3 | 11450 | NA | NA | NA | NA | NA | NA | 39/61 |
| scalefree | 4174 | 5.7 | 11792 | 1 | 1 | NA | NA | NA | NA | 41/59 |
| smallworld | 6869 | 4.0 | 13582 | 1 | 1 | NA | NA | NA | NA | 42/58 |
| Protein | 4626 | 6.4 | 14801 | 7 | 12 | 3434 | 6 | 23 | 1 | 36/64 |
| random | 4620 | 6.3 | 14653 | NA | NA | NA | NA | NA | NA | 43/57 |
| scalefree | 3459 | 8.5 | 14771 | 2 | 16 | 2828 | 4 | 14 | 1 | 36/64 |
| smallworld | 4623 | 6.0 | 13878 | 1 | NA | NA | NA | NA | NA | 44/56 |
| Internet | 9200 | 6.3 | 28957 | 7 | 13 | 8103 | 6 | 117 | 1 | 35/65 |
| random | 9182 | 6.3 | 28864 | NA | NA | NA | NA | NA | NA | 43/57 |
| scalefree | 6905 | 8.5 | 29336 | 2 | 13 | 5400 | 7 | 29 | 1 | 34/66 |
| smallworld | 9188 | 6.0 | 27600 | 1 | NA | NA | NA | NA | NA | 44/56 |
| Streets | 4501 | 5.1 | 11408 | 5 | 10 | 2020 | 6 | 30 | 1 | 17/83 |
| random | 4463 | 5.0 | 11167 | NA | NA | NA | NA | NA | NA | 41/59 |
| scalefree | 3161 | 7.3 | 11153 | 8 | 11 | 2525 | 6 | 23 | 1 | 36/64 |
| smallworld | 4497 | 6.0 | 13503 | 1 | NA | NA | NA | NA | NA | 44/56 |
| WWW | 213 | 122.0 | 12994 | 4 | 1 | 6 | 2 | 153 | 1 | 1/99 |
| random | 213 | 122.0 | 13005 | NA | NA | NA | NA | NA | NA | 48/52 |
| scalefree | 206 | 116.2 | 11966 | 8 | 13 | 180 | 3 | 15 | 1 | 25/75 |
| smallworld | 213 | 112.0 | 12993 | 1 | NA | NA | NA | NA | NA | 48/52 |

Before discussing the experimental results, we briefly introduce the ht-index for characterizing heavy-tailed distributions or the scaling pattern of far more small things than large ones (Jiang and Yin 2014). The ht-index was developed as an alternative to fractal dimension for quantifying the complexity of fractals: the higher the ht-index, the more complex the fractals. The ht-index is h if the scaling pattern of far more small things than large ones recurs h-1 times. For example, the edge betweenness of the fictive network has an ht-index of 3 because the scaling pattern recurs twice. Panel D of Figure 1 shows the first occurrence: seven in the head and 13 in the tail; and the seven edges in the head are further partitioned around the second mean of 0.191, two in the head and five in the tail, indicating the second occurrence. In the following experiments, the ht-index is used to indicate heavy-tailed distributions for both edge betweenness and community size.

For each network, we generate three counterparts – random (Erdős and Rényi 1959), scale-free (Barabási and Albert 1999), and small-world graphs (Watts and Strogatz 1998) – using the same numbers of nodes and edges (note that, eventually, the numbers of nodes and edges are approximately the same because of isolated nodes that are excluded). The random graph is used to create the reference head/tail ratio to determine whether networks or their subnetworks are homogeneous enough. The scale-free and small-world networks are used for comparison purposes. For most networks, their equivalent small-world networks contain no communities except for Brightkite and Gowalla, but their scale-free networks contain communities except for Erdös and Gowalla (Table 1). The resulting communities demonstrate the scaling property: far more small communities than large ones, because the ht-index is between 3 and 8. For example, Brightkite is broken down into 172 homogeneous communities with sizes between 27 and 1 (Table 1). Because the scaling pattern of far more small communities than large ones recurs twice, the ht-index is 3. Figure 3 provides a sense of how the network is broken into pieces in a step-by-step manner. Before the tenth iteration, many heterogeneous communities were generated in each step. Panel F of Figure 3 shows a rank-size plot for all communities, both heterogeneous and homogeneous, generated from different iterations. Strikingly, community sizes become increasingly scaling as iteration increases.

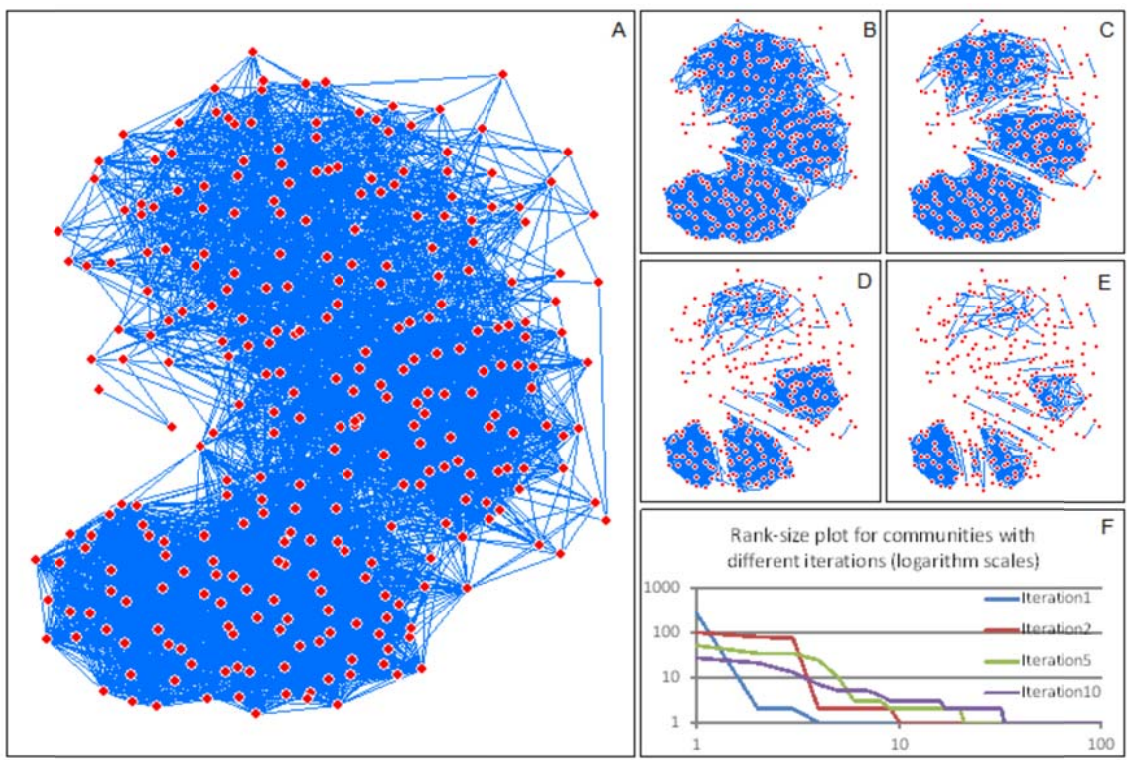

Figure 3: (Color online) Head/tail breaking the Brightkite network into communities
(Note: the original network (Panel A) is broken into many homogeneous communities after 10 iterations, Panels B, C, D, and E represent iterations 1, 2, 5, and 10, Panel F shows how the communities become increasingly scaling as the iteration increases.)

The experiments generated a random graph for each network to determine the reference head/tail ratio or the head percentage. There is probably no need to do so because the head percentage is always less than 50 for all of the networks. Therefore, we can select a fixed head/tail ratio, such as 45/55, 40/60, and 50/50. Table 2 presents the results for how the derived communities (both heterogeneous and homogeneous) demonstrate heavy-tailed distributions characterized by the ht-index and power laws characterized by alpha and p using the robust maximum likelihood estimation (Clauset et al. 2009). This part of the experiments indicates that a head/tail ratio of 45/55 appears to be a good approximation of all of the random graphs. Therefore, we suggest a head percentage of 45 for any complex network when applying the community detection algorithm to achieve reasonably good results. Note that in Table 2, the number of communities for the WWW network is only six, which is too small to obtain statistically significant results for the power law detection.

Table 2: Heavy-tailed distributions or power laws of all communities with different head/tail ratios (Note: # acom = number of all communities, ht-acom = ht-index for all communities, alpha = power law exponent for all communities, p = index indicating goodness-of-fit for power laws (Clauset et al. 2009), h/t = head/tail ratio of the corresponding random graphs with respect to their edge betweenness, NA = not available. Table 2 is divided into four panels, with Panel A using the random graph head/tail ratios and Panels B, C, and D using the fixed ratios of 45/55, 40/60, and 50/50.)

| A | # acom | ht-acom | alpha | p | h/t | B | # acom | ht-acom | alpha | p | h/t |
|---|---|---|---|---|---|---|---|---|---|---|---|
| Brightkite | 206 | 5 | 2.21 | 0.35 | 48/52 | | 207 | 5 | 2.21 | 0.31 | 45/55 |
| Gowalla | 453 | 4 | 1.92 | 0.12 | 49/51 | | 424 | 4 | 1.75 | 0.13 | 45/55 |
| Scientific | 7222 | 5 | 2.82 | 0.11 | 43/57 | | 7824 | 5 | 2.85 | 0.11 | 45/55 |
| Erdös | 3262 | 5 | 2.69 | 0.16 | 39/61 | | 3532 | 5 | 2.72 | 0.15 | 45/55 |
| Protein | 3767 | 5 | 2.45 | 0.03 | 43/57 | | 3806 | 5 | 2.43 | 0.03 | 45/55 |
| Internet | 8398 | 5 | 3.3 | 0.05 | 43/57 | | 8462 | 5 | 3.24 | 0.04 | 45/55 |
| Streets | 2642 | 4 | 2.05 | 0.16 | 41/59 | | 3027 | 4 | 2.09 | 0.15 | 45/55 |
| WWW | 6 | 2 | NA | NA | 48/52 | | 6 | 2 | NA | NA | 45/55 |
| C | | | | | | D | | | | | |
| Brightkite | 192 | 4 | 2.26 | 0.28 | 40/60 | | 207 | 5 | 2.21 | 0.31 | 50/50 |
| Gowalla | 361 | 4 | 1.66 | 0.16 | 40/60 | | 455 | 6 | 1.91 | 0.1 | 50/50 |
| Scientific | 6091 | 5 | 2.76 | 0.09 | 40/60 | | 8478 | 6 | 2.76 | 0.11 | 50/50 |
| Erdös | 3523 | 5 | 2.72 | 0.16 | 40/60 | | 3532 | 5 | 2.72 | 0.15 | 50/50 |
| Protein | 1472 | 3 | 4.15 | 0 | 40/60 | | 3958 | 5 | 2.36 | 0.03 | 50/50 |
| Internet | 2564 | 3 | 5.59 | 0 | 40/60 | | 8586 | 6 | 3.21 | 0.22 | 50/50 |
| Streets | 2508 | 4 | 2.04 | 0.17 | 40/60 | | 3216 | 6 | 2.11 | 0.23 | 50/50 |
| WWW | 6 | 2 | NA | NA | 40/60 | | 6 | 2 | NA | NA | 50/50 |

Table 3: The communities generated by the edge removing algorithm in comparison with ours (Note: Sum = total number of communities, Min = minimum size of communities, Max = maximum size of communities, %Match = percentage match with the communities by our method, Q = highest modularity)

| | Sum | Min | Max | Mean | %Match | Q |
|---|---|---|---|---|---|---|
| Fictive | 3 | 3 | 5 | 4.0 | 67% | 0.47 |
| Club | 5 | 2 | 13 | 7.8 | 40% | 0.44 |
| Brightkite | 16 | 2 | 103 | 19.0 | 25% | 0.46 |
| Gowalla | 255 | 2 | 216 | 3.1 | 31% | 0.12 |
| Scientific | 159 | 6 | 1009 | 81.0 | 6% | 0.81 |
| Erdös | 100 | 4 | 931 | 70.3 | 17% | 0.70 |
| Protein | 347 | 3 | 741 | 14.3 | 2% | 0.46 |
| Internet | 223 | 3 | 1454 | 42.3 | 0% | 0.40 |
| Streets | 17 | 39 | 504 | 265.8 | 0% | 0.76 |
| WWW | 4 | 5 | 155 | 54.3 | 50% | 0.14 |

We also compared our method to the edge removing algorithm (Girvan and Newman 2002), which removes high betweenness edges step by step, until none is left. In the edge removing process, it is suggested to calculate the modularity value each time when a new component (or community) has emerged (Newman 2004). Eventually, the outcomes with the highest modularity (Q in Table 3) are considered to be the detected communities. We examined how many of communities by the edge removing algorithm could find their counterparts in our method-induced communities (Table 3). The table shows that all the networks have a very low match percentage of communities. This suggests that our method is very unique, and its results cannot be compared to those of the previous method. The comparison further reinforces our belief that unlike simple networks, self-organized and/or self–evolved networks, or complex networks in general, cannot be easily decomposed into parts, for the parts tend to mutually entangled. On the one hand, they are nested, corresponding to our heterogeneous and homogeneous communities; on the other hand, they tend to be very heterogeneous in sizes. In this connection, simple networks are very much like mechanical watches that are decomposable, while complex networks like human brains that are hard to decompose. This is the fundamental thinking that differentiates our method from previous methods.

### IV. Further discussions and conclusion

This work is very much inspired by the natural cities extracted from social media location data (Jiang and Miao 2014, Jiang 2015). Individual users' check-in locations constitute a large triangle irregular network (TIN) whose edges demonstrate a heavy-tailed distribution, i.e., far more short edges than long ones. Eventually, all short edges (shorter than an average of all the edges) constitute different clumps called natural cities. In a similar manner, there are far more small betweenness edges than large ones for complex networks, indicating that they contain many clumps called communities. The major differences between the natural cities derived from the TIN and the communities from complex networks are as follows: (1) the TIN is partitioned only once to obtain the natural cities, whereas a complex network is partitioned multiple times recursively to obtain communities; therefore, (2) the derived communities are nested, whereas the natural cities are not. However, for the natural cities, we can also recursively continue the partition process to obtain hotspots in the cities. This way the natural cities and hotspots (both as communities) would be nested as well. The nested relationships are frequently seen in reality, e.g., a country as a set of cities, a city as a set of neighborhoods, and a neighborhood as a set of families. One disadvantage of the community detection algorithm lies in the computational complexity of the edge betweenness, in particular for large networks. In our experiments, we were able to afford to use only parts of some large networks, such as Brightkite, Gowalla, and WWW.

Previous studies relied on real-world networks with known communities to verify community detection algorithms. This verification approach is questionable because the known communities could still be very heterogeneous and should be further partitioned into homogeneous ones. For example, the club network contains two known communities (Zachary 1977, Girvan and Newman 2002), whereas our algorithm leads to three heterogeneous communities and 11 homogeneous ones. Intuitively, the fictive network contains three communities; instead, our algorithm results in four communities. The reader may ask how to verify our results. We believe that the scaling pattern of far more small things than large ones is universal and applies to the communities of a network as well if the network is self-organized and/or naturally evolved. We further believe that the community detection process leading to far more small communities than large ones is very similar to dropping a piece of glass into stone, resulting in far more small pieces than large ones – the fractal or scaling nature of the broken pieces. In other words, we use the scaling pattern to verify our results.

The notion behind the community detection algorithm is holistic, i.e., taking all edges as a whole and classifying them into the head and tail or, equivalently, the outside and inside edges, and recursively continuing the classification for the inside edges until a network and its subnetworks become homogeneous enough. From the holistic perspective, whether a family is a community is relative to the other families to which it links and to the random graph counterpart. Surprisingly, we found that the derived communities demonstrate a striking scaling property, i.e., far more small homogeneous communities than large ones. During the iterative partitioning, many heterogeneous or large communities can be identified at different coarse-graining levels. The scaling property is even more striking by taking both homogeneous and heterogeneous together, and this is shown by power laws of the communities for all large networks.

**Acknowledgments**
We would like to thank the editor and two anonymous reviewers, in particular the second, who provide some valuable comments that have significantly improved our work.